\documentclass[prl,aps,twocolumn,showpacs,citeautoscript,reprint,superscriptaddress]{revtex4-2}
\usepackage{graphicx}
\usepackage{bm,amsmath,amssymb,wasysym,amsfonts,dcolumn}
\usepackage[colorlinks=true, urlcolor=blue, linkcolor=blue, citecolor=blue, pdfborder={0 0 0}]{hyperref}
\usepackage[usenames,dvipsnames]{xcolor}
\usepackage{mathrsfs}

\begin{document}

\title{Fourfold anisotropic magnetoresistance of L1$_0$ FePt due to relaxation time anisotropy}

\author{Y. Dai}
\altaffiliation{These authors contributed equally to this study.}
\affiliation{Shanghai Key Laboratory of Special Artificial Microstructure Materials and Technology and Pohl Institute of Solid State Physics and School of Physics Science and Engineering, Tongji University, Shanghai 200092, China}
\author{Y. W. Zhao}
\altaffiliation{These authors contributed equally to this study.}
\affiliation{Center for Advanced Quantum Studies and Department of Physics, Beijing Normal University, Beijing 100875, China}
\author{L. Ma}
\author{M. Tang}
\author{X. P. Qiu}
\affiliation{Shanghai Key Laboratory of Special Artificial Microstructure Materials and Technology and Pohl Institute of Solid State Physics and School of Physics Science and Engineering, Tongji University, Shanghai 200092, China}
\author{Y. Liu}
\author{Z. Yuan}
\email{zyuan@bnu.edu.cn}
\affiliation{Center for Advanced Quantum Studies and Department of Physics, Beijing Normal University, Beijing 100875, China}
\author{S. M. Zhou}
\email{shiming@tongji.edu.cn}
\affiliation{Shanghai Key Laboratory of Special Artificial Microstructure Materials and Technology and Pohl Institute of Solid State Physics and School of Physics Science and Engineering, Tongji University, Shanghai 200092, China}
\date{\today}

\begin{abstract}

Experimental measurements show that the angular dependence of the anisotropic magnetoresistance (AMR) in L1$_0$ ordered FePt epitaxial films on the current orientation and magnetization direction is a superposition of the corresponding dependences of twofold and fourfold symmetries. The twofold AMR exhibits a strong dependence on the current orientation, whereas the fourfold term only depends on the magnetization direction in the crystal and is independent of the current orientation. First-principles calculations reveal that the fourfold AMR arises from the relaxation time anisotropy due to the variation of the density of states near the Fermi energy under rotation of the magnetization. This relaxation time anisotropy is a universal property in ferromagnetic metals and determines other anisotropic physical properties that are observable in experiment.
\end{abstract}

\maketitle

{\it\color{red}Introduction.---}The fundamental physics of spintronics is the interplay of magnetization in magnetic materials and electrical currents~\cite{spintronics}. The electrical resistance of a magnetic device typically depends on the magnetization configuration, resulting in a variety of intriguing magnetoresistance (MR) phenomena, such as spin-Hall MR~\cite{Nakayama2013}, Rashba-Edelstein MR~\cite{Nakayama2016}, spin-orbital MR~\cite{Zhou2018}, anomalous Hall MR~\cite{Yang2018}, and Hanle MR~\cite{Velez2016}, which are effectively applied in probing a magnetic field or magnetization. As a basic MR effect in ferromagnetic metals (FMs) and alloys, AMR describes the dependence of electrical resistivity on the magnetization direction~\cite{Thomson1857,Campbell1923,Potter1974,Hupfauer2015,Nadvornik2021,Park2021}. 

AMR and its angular dependence on the current orientation and magnetization direction are attributed to the interaction among the crystal field, exchange field and spin-orbit coupling (SOC)~\cite{Berger1968,Campbell1970,Kokado2012}. Most early studies were carried out on polycrystalline samples, in which the symmetry constraint only allowed a twofold term that scaled as $\cos^2\varphi_M$, where $\varphi_M$ is the angle between the magnetization and current~\cite{McGuire1975}. Twofold AMR has multiple microscopic mechanisms including $s$-$d$ scattering~\cite{Campbell1970,Trushin2009}, the intrinsic mechanism from band crossings~\cite{Zeng2020} and band splitting due to lattice distortion~\cite{Kokado2015}. Fourfold AMR has been experimentally observed in Fe, Co and Ni epitaxial films~\cite{Gorkom2001,Xiao2015,Zeng2020a,Ye2013,Xiao2015a}, (Ga,Mn)As~\cite{Limmer2006,Rushforth2007}, manganites~\cite{Bason2009,Naftalis2009}, Fe$_3$O$_4$~\cite{Naftalis2011,Ding2013}, Co$_2$MnSi~\cite{Oogane2018} and antiferromagnetic EuTiO$_3$~\cite{Ahadi2019}. It exists in pseudoepitaxial Fe$_4$N films~\cite{Tsunoda2010,Li2015,Kokado2015} at low temperatures and vanishes at elevated temperatures. The latter was ascribed to the tetragonal lattice distortion in Fe$_4$N~\cite{Kokado2015}, but this interpretation is not applicable to cubic Ni~\cite{Xiao2015}. Recently, the fourth-order perturbation of SOC was proposed to be the mechanism in cubic crystals~\cite{Yahagi2020}. So far, fourfold AMR is still poorly understood.

L1$_0$ Fe$_{0.5}$(Pd$_{1-x}$Pt$_{x}$)$_{0.5}$ is an ordered ferromagnetic alloy in which both the degree of chemical ordering and SOC strength are tunable and is therefore an ideal material to investigate the microscopic mechanisms underlying AMR. Systematic measurement of the resistivity of FePt epitaxial films combined with first-principles calculations allows us to gain a thorough understanding of the observed angular dependence of AMR. Using the current-orientation independence as the criterion, we discover that the fourfold AMR arises from the variation in the density of states near the Fermi surface, which results in the relaxation time anisotropy under rotation of the magnetization with respect to the crystallographic axes. 

\begin{figure}[t]
  \centering
  \includegraphics[width=\columnwidth]{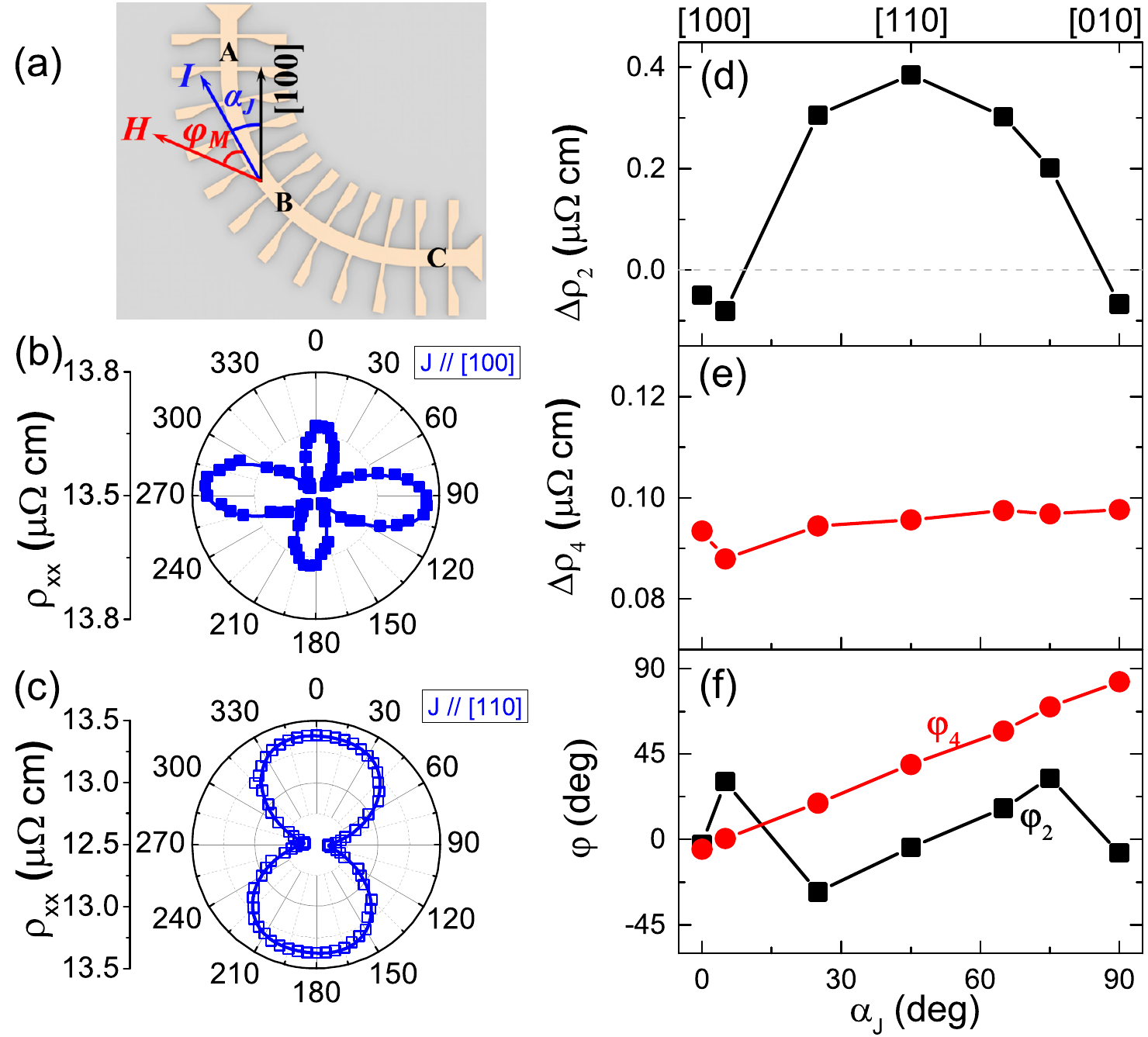}
  \caption{(a) Schematic of AMR measurement. The measured resistivity $\rho_{xx}$ of the L1$_0$ FePt film with $S=0.82$ at point A ($\mathbf J\|[100]$) and point B ($\mathbf J\|[110]$) is plotted in (b) and (c), respectively, as a function of the magnetization direction ($\varphi_M$) with respect to the sensing current. The solid lines in (b) and (c) correspond to fits using Eq. (\ref{eq:1}). The fitted parameters $\Delta\rho_{2(4)}$ and $\varphi_{2(4)}$ are shown in (d), (e) and (f) as a function of the current orientation ($\alpha_{\mathbf J}$) with respect to the crystalline axis [100]. The measurement is carried out at a low temperature of $T=10$~K.}
  \label{fig1}
\end{figure}

{\it\color{red}Measured AMR of FePt.---}
A single-crystal FePt (001) film is epitaxially grown on an MgO (001) substrate and patterned into arcuate Hall bars so that the current direction is continuously variable, as schematized in Fig.~\ref{fig1}(a). The degree of chemical ordering $S$ is controlled by the substrate temperature during fabrication and the postannealing temperature~\cite{SI}. $S=1$ for a fully ordered structure, and $S=0$ for a completely disordered alloy. 

The magnetization within the (001) plane is rotated to measure the longitudinal resistivity $\rho_{xx}$ of FePt with $S=0.82$ for currents along [100] and [110], as shown in Fig.~\ref{fig1}(b) and (c), respectively. The measured data are effectively fitted using the following superposition of the twofold and fourfold AMR terms:
\begin{eqnarray}
\rho_{xx}(\varphi_M)&=&\rho_0+\Delta\rho_2\cos2(\varphi_{M}+\varphi_{2})\nonumber\\
&&+\Delta\rho_4\cos4(\varphi_M+\varphi_4),\label{eq:1}
\end{eqnarray}
where $\varphi_M$ represents the angle between the magnetization and sensing current defined in Fig.~\ref{fig1}(a) and $\rho_0$ is the average resistivity independent of $\varphi_M$. The last two terms in Eq.~\eqref{eq:1} correspond to the twofold and fourfold variations in resistivity, with phases of $\varphi_2$ and $\varphi_4$, respectively. In Fig.~\ref{fig1}(d) and (e), the fitted $\Delta\rho_2$ and $\Delta\rho_4$ are plotted as a function of the current direction $\alpha_{\mathbf J}$, which is defined by the angle between the current direction and the crystal axis [100] [see Fig.~\ref{fig1}(a)]. $\Delta\rho_2$ is highly sensitive to the sensing current direction and  is small and negative at $\mathbf J\|[100]$ and becomes large and positive at $\mathbf J\|[110]$. By contrast, the fourfold term $\Delta\rho_4$ has a nearly constant magnitude that varies by less than 10\% over the range of $0\le\alpha_{\mathbf J}\le90^\circ$. The associated phase $\varphi_4$ in the fourfold term is always equal to the current orientation $\alpha_{\mathbf J}$, as shown in Fig.~\ref{fig1}(f). This result suggests that fourfold AMR is independent of the current orientation and depends only on the magnetization direction with respect to the crystallographic axes. Unlike $\varphi_4$, the phase $\varphi_2$ in the twofold term exhibits nonmonotonic variation between $-45^\circ$ and $45^\circ$ suggesting competition of multiple components~\cite{SI}.

\begin{figure}[t]
  \centering
  \includegraphics[width=\columnwidth]{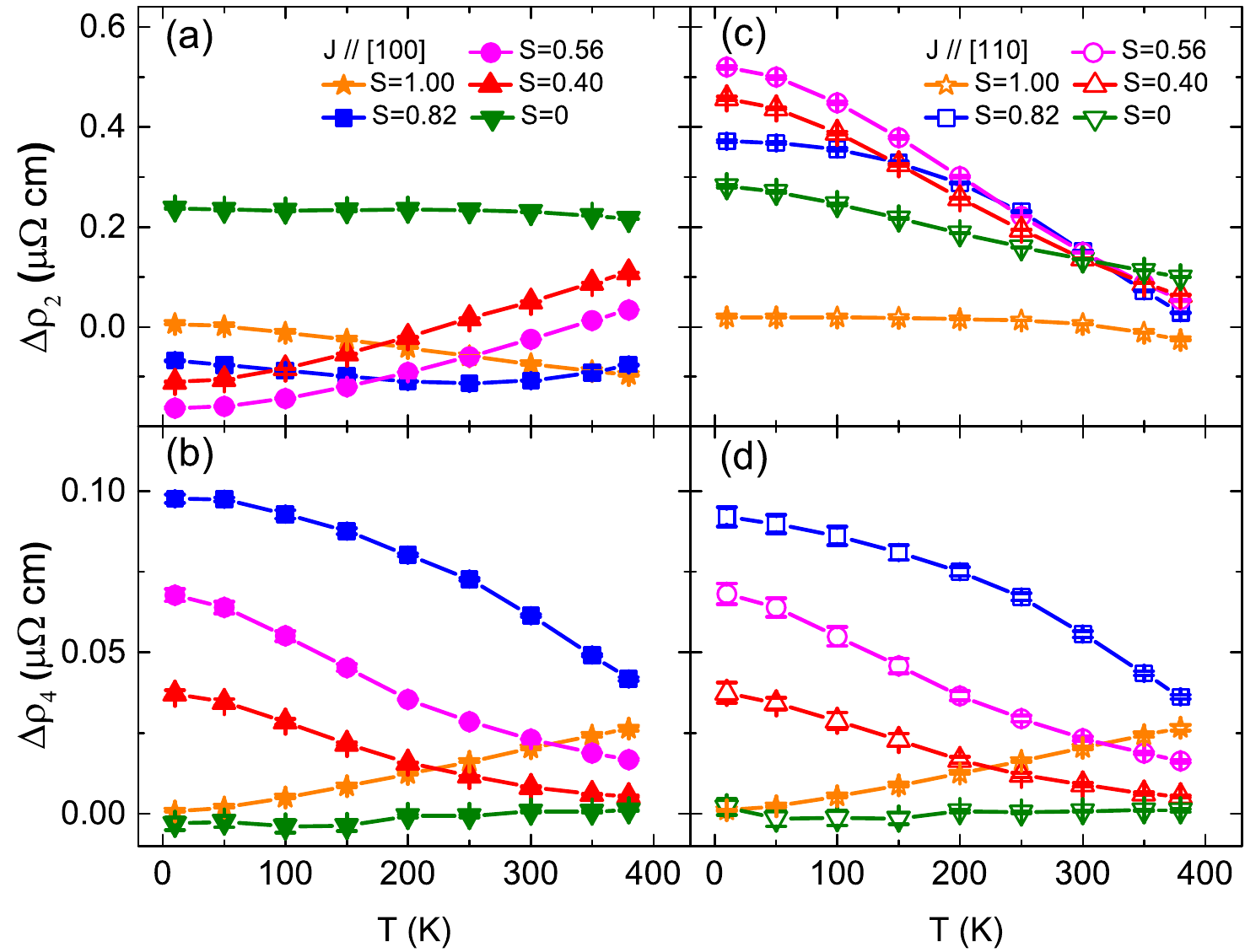}
  \caption{Fitted AMR parameters $\Delta\rho_2$ and $\Delta\rho_4$ as a function of temperature for samples with different degrees of chemical order. The measurement is carried out under a fixed current orientation that is along [100] in (a) and (b) and along [110] in (c) and (d).}
  \label{fig2}
\end{figure}
Figure~\ref{fig2} shows $\Delta\rho_2$ and $\Delta\rho_4$ that are extracted using Eq.~\eqref{eq:1} from experimental data~\cite{SI} as a function of the temperature for samples with various degrees of chemical order. Here, we focus on two sensing current directions along high-symmetry axes, [100] and [110]. For $\mathbf J\|[100]$, $\Delta\rho_2$ is negative at large $S$ [see the orange and blue symbols in Fig.~\ref{fig2}(a)]. At intermediate $S$, the twofold AMR exhibits a transition from negative at low temperatures to positive at room temperature and above. In the completely disordered sample, $\Delta\rho_2$ is always positive and nearly invariant with increasing temperature. The temperature dependence of $\Delta\rho_2$ is strikingly different for $\mathbf J\|[110]$ in Fig.~\ref{fig2}(c), where it decreases with increasing temperature for all samples. Comparing Fig.~\ref{fig2}(b) and (d), we find that the fourfold AMR has the same temperature dependence for $\mathbf J\|[100]$ and $\mathbf J\|[110]$. With increasing temperature, $\Delta\rho_4$ increases for $S=1$ and decreases for smaller $S$. The fourfold AMR vanishes in the sample with $S=0$.

\begin{figure}[t]
  \centering
  \includegraphics[width=\columnwidth]{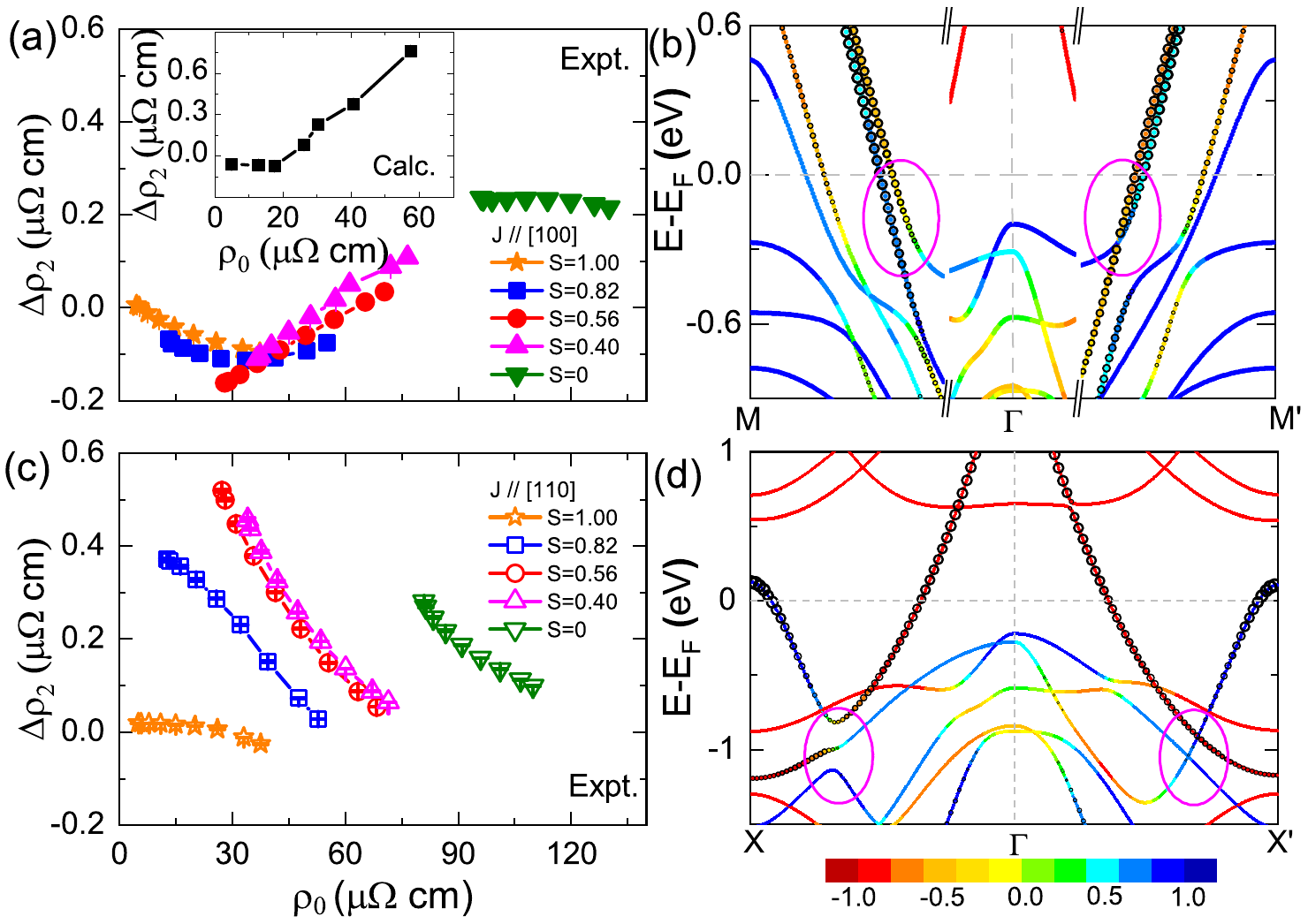}
  \caption{Fitted $\Delta\rho_2$ as a function of the average resistivity $\rho_{0}$ for current along [100] (a) and [110] (c). Inset of (a): Calculated $\Delta\rho_2$ as a function of $\rho_{0}$ for fully-ordered FePt with thermal lattice disorder. Calculated band structure with SOC along $\langle100\rangle$ (b) and along $\langle110\rangle$ (d). The magnetization is parallel to $\Gamma\mathrm M'$ ($\Gamma\mathrm X'$) and perpendicular to $\Gamma\mathrm M$ ($\Gamma\mathrm X$). The colors of the energy bands indicate the spin projection along the quantization axis. The circles with different sizes represent the $sp$ components of the Bloch states.}
  \label{fig3}
\end{figure}

{\it\color{red}Twofold AMR.---}It is difficult to conclusively determine the AMR dependence on the degree of chemical ordering and temperature from Fig.~\ref{fig2} because both factors influence the AMR simultaneously. To obtain deeper insight into this dependence, we replot the measured $\Delta\rho_2$ of all the samples as a function of the corresponding average resistivity $\rho_0$ for $\mathbf J\|[100]$ in Fig.~\ref{fig3}(a). A common trend in the experimental data is thus revealed: $\Delta\rho_2$ is negative at low resistivities and becomes positive at large $\rho_0$. For the fully disordered alloy with $S=0$, $\Delta\rho_2$ is a positive constant that is independent of $\rho_0$. 

We perform a first-principles transport calculation for fully-ordered L1$_0$ FePt with $S=1$, where temperature-induced lattice disorder is introduced to account for the finite resistivity~\cite{SI, Starikov2018}. By increasing the temperature, we qualitatively reproduce the resistivity dependence of twofold AMR, as shown in the inset of Fig.~\ref{fig3}(a), and the difference at small $\rho_0$ is attributed to the large perpendicular anisotropy of highly ordered FePt~\cite{SI}. The negative $\Delta\rho_2$ at low $\rho_0$ and positive value at large $\rho_0$ can be understood by analyzing the band structure of L1$_0$ FePt. Figure~\ref{fig3}(b) shows the energy bands near the Fermi energy $E_F$ along the high-symmetry direction $\langle100\rangle$ for parallel ($\Gamma \mathrm M'$) or perpendicular ($\Gamma \mathrm M$) magnetizations. In the low disorder regime, the twofold AMR is nearly independent of the scattering rate or relaxation time, suggesting an intrinsic contribution due to band (anti)crossing~\cite{Zeng2020}. These special band crossings are a consequence of symmetry at a given $\mathbf M$, whereas rotating $\mathbf M$ breaks the symmetry and lifts the band degeneracy. As the energy bands near $E_F$ have the characteristics of $sp$-$d$ hybridization, we focus on the itinerant $sp$ bands (marked by empty circles) that have a stronger influence on transport than the more localized $d$ bands. Rotating $\mathbf M$ from [100] to [010] results in the disappearance of a crossing of $sp$ bands along [100] ($\Gamma \mathrm M'$) (marked by purple ellipses) and hence slightly increases the resistivity, corresponding to a negative $\Delta\rho_2$. As we increase $\rho_0$ by increasing the temperature or decreasing $S$, the contribution of disorder scattering to the AMR becomes more important. Most of the $d$ bands near $E_F$ have the minority-spin component~\cite{Khan2016}, and therefore, $sp$-$d_{\downarrow}$ scattering leads to a positive $\Delta\rho_2$~\cite{McGuire1975}.

For $\mathbf J\|[110]$, an arbitrary $\rho_0$ corresponds to different $\Delta\rho_2$ depending on the details of the degree of chemical order; see Fig.~\ref{fig3}(c). This behavior implies that the AMR mainly arises from extrinsic disorder scattering, in agreement with the calculated band structure. As shown in Fig.~\ref{fig3}(d), the energy bands along [110] near $E_F$ are unchanged for $\mathbf M\|[110]$ ($\Gamma \mathrm X'$) or $\mathbf M\|[1\bar{1}0]$ ($\Gamma \mathrm X$) and hence make no intrinsic contribution to the AMR. When disorder scattering becomes stronger with increasing temperature, the resulting band smearing enables the energy bands farther away from $E_F$ to affect electron transport. The band crossings along $\Gamma\mathrm X'$ become anticrossing gaps along $\Gamma\mathrm X$ (marked by purple ellipses), corresponding to a negative $\Delta\rho_2$ from the intrinsic mechanism. Therefore, the experimentally measured $\Delta\rho_2$ for all $S$ decreases with increasing temperature.

\begin{figure}[b]
  \centering
  \includegraphics[width=\columnwidth]{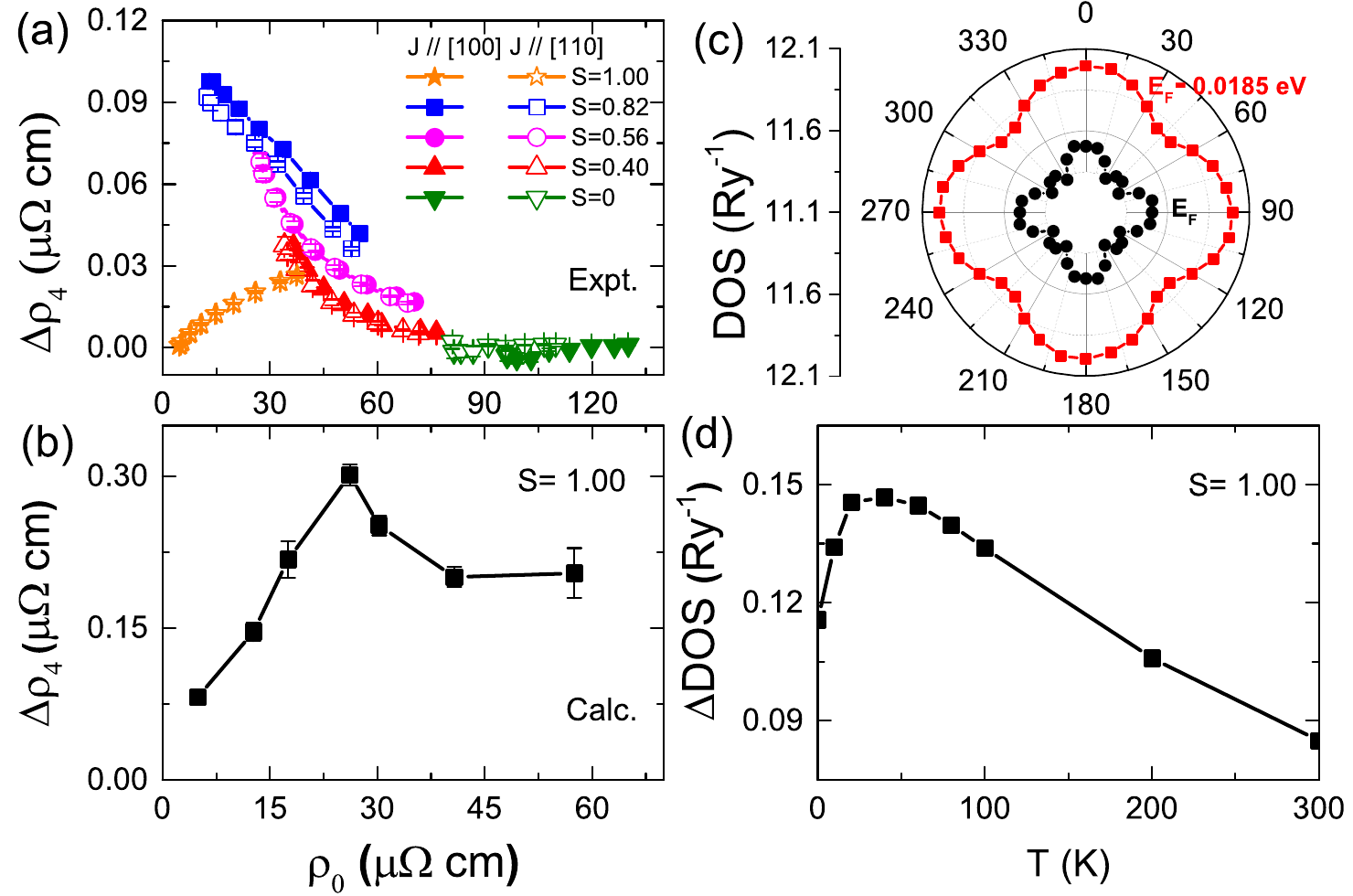}
  \caption{(a) Fitted $\Delta\rho_4$ from experimentally measured resistivity. (b) Calculated $\Delta\rho_4$ as a function of the total resistivity for fully-ordered FePt. (c) Calculated DOS of FePt as a function of the magnetization direction rotated within the (001) plane. The black circles and red squares are calculated at $E_F$ and $E_F-0.0185$~eV, respectively. The Brillouin zone is sampled by $\sim 2400^3$ $k$-points for convergence. (d) The difference in the finite-temperature DOS for $\mathbf M\|[100]$ and $\mathbf M\|[110]$ calculated using Eq.~\eqref{eq:3}.}
  \label{fig4}
\end{figure}
{\it\color{red}Fourfold AMR.---}The experimental fourfold AMR $\Delta\rho_4$ is plotted in Fig.~\ref{fig4}(a) as a function of the measured average resistivity $\rho_0$. The data extracted from different sensing current directions overlap with each other. With increasing $\rho_0$, $\Delta\rho_4$ of the highly ordered samples increases and it decreases for samples with smaller $S$. The strong dependence of $\Delta\rho_4$ on the ordering degree and temperature indicates that the physical mechanism for fourfold AMR is related to scattering. Moreover, the measurements with various sample thicknesses confirm that fourfold AMR is not a surface effect but exists in bulk L1$_0$ FePt~\cite{SI}. A first-principles transport calculation of fully-ordered L1$_0$ FePt also reproduces the nonmonotonic $\rho_0$-dependent fourfold AMR: with increasing temperature-induced thermal lattice disorder, the calculated $\Delta\rho_4$ increases to a maximum and then decreases, as shown in Fig.~\ref{fig4}(b). In addition, the calculated resistivity also exhibits maxima and minima at $\mathbf M\|\langle100\rangle$ and $\mathbf M\|\langle110\rangle$, respectively~\cite{SI}, in agreement with experiment. The calculated $\Delta\rho_4$ is much larger than the experimental values and this discrepancy may be attributed to the neglected spin fluctuation and chemical disorder in the calculation.

Fourfold AMR in Fe$_4$N is attributed to energy-band hybridization resulting from the interplay of SOC and tetragonal distortion~\cite{Kokado2015}. The predictions of this theory, however, contradict both our experimental observations and calculations for  $\Delta\rho_4$. Experimentally, the structural L1$_0$-A$_1$ phase transition of FePt occurs near $1300^{\circ}$C ~\cite{Massalski1986}, below which the tetragonal structure of FePt films is sustained. However, the measured $\Delta\rho_4$ becomes very small near 400~K far below the phase transition temperature, as seen in Fig.~\ref{fig2}(b) and (d), especially for small $S$. In our calculation, the tetragonal structure of L1$_0$ FePt is not affected by temperature, whereas the calculated $\Delta\rho_4$ exhibits a nonmonotonic dependence. 

Under rotation of the magnetization, SOC mediates the electronic states in the crystal field, leading to a variation in the density of states (DOS) at the Fermi energy. Using a hybrid Wannier-Bloch representation~\cite{Yuan2016}, we calculate the DOS of fully-ordered L1$_0$ FePt and two typical cases at $E_F$ and $E_F-0.0185$~eV are plotted in Fig.~\ref{fig4}(c), both of which show fourfold symmetry and the maxima (minima) at $\mathbf M\|\langle100\rangle$ ($\mathbf M\|\langle110\rangle$). The electronic scattering rate at the Fermi level is inversely proportional to the relaxation time and determined by the Fermi golden rule~\cite{Ziman1960}, 
\begin{equation}
\frac{1}{\tau}\propto\frac{2\pi}{\hbar}\vert\langle f\vert V\vert i\rangle\vert^2 D_f D_i.\label{eq:2}
\end{equation}
Here, we only consider the dominant elastic scattering contribution due to the scattering potential $V$. In Eq.~\eqref{eq:2}, $\vert i\rangle$ and $\vert f\rangle$ are the initial and final states, respectively. $D_i$ and $D_f$ represent the densities of these states at the Fermi energy. For $\mathbf M\|\langle100\rangle$, the increase in the DOS enhances the scattering probability of Bloch states and thus reduces the relaxation time. This picture explains why the fourfold AMR only depends on the magnetization direction with respect to the crystallographic axes and is invariant for different sensing current directions. 

With increasing chemical and lattice disorder, energy-band smearing causes states away from $E_F$ to be incorporated into the DOS at the Fermi energy. By introducing a Fermi-Dirac distribution function at finite temperature $f(E,T)$, we calculate the DOS at $E_F$ as 
\begin{equation}
D(E_F, T)=\int_{-\infty}^{\infty}dE\,D(E)\left[-\frac{\partial f(E,T)}{\partial E}\right].\label{eq:3}
\end{equation}
The difference of calculated DOS between $\mathbf M\|\langle100\rangle$ and $\mathbf M\|\langle110\rangle$ is plotted in Fig.~\ref{fig4}(d). The anisotropic DOS also exhibits a nonmonotonic dependence on temperature that is consistent with that of the fourfold AMR. This nonmonotonic behavior is attributed to the fact that the DOS has a larger anisotropy at $E_F+\epsilon$ than at $E_F$. Thus, increasing the temperature enables the energy $E_F+\epsilon$ to contribute to transport and enhances the fourfold AMR. An enough high temperature involves a very large energy range and eventually averages out the anisotropy in the DOS. Note that the calculated $\Delta\rho_4$ in Fig.~\ref{fig4}(b) and anisotropic DOS in Fig.~\ref{fig4}(d) decrease more slowly with the temperature than the experimental $\Delta\rho_4$ in Fig.~\ref{fig4}(a). This result is obtained because chemical disorder and spin wave excitations, which are not included in the calculation, strongly suppress the fourfold AMR in reality by breaking local symmetry. 

With SOC that couples spin and real space, the particular magnetization orientation affects the electronic structure and thus modulates the DOS near Fermi surface. Such a modulation follows the crystal symmetry and hence leads to fourfold AMR in any metallic ferromagnets with fourfold symmetry. By carefully studying the literature, we have found that the measured fourfold AMR in Co~\cite{Xiao2015}, Ni~\cite{Xiao2015a} and Fe$_4$N~\cite{Tsunoda2010} are independent of current direction in agreement with our findings. It indicates that the proposed microscopic mechanism for the fourfold AMR is universal for ferromagnetic metals.

In addition to resistivity, relaxation time is important for Gilbert damping that characterizes dynamical magnetization dissipation~\cite{Gilbert2004}. At low temperature, Kambersk{\'y}'s breathing Fermi surface model~\cite{Kambersky1970} explicitly shows the proportionality of Gilbert damping and relaxation time, while Gilbert damping at high temperature is inversely proportional to relaxation time due to the dominant interband scattering~\cite{Kambersky2007}. This nonmonotonic relationshop has been demonstrated by first-principles calculations~\cite{Gilmore2007}. Thus the relaxation time anisotropy due to the modulation of DOS shall lead to anisotropic Gilbert damping~\cite{Gilmore2010}, which was recently observed in single-crystal ferromagnets but has not yet been understood~\cite{Li2019,Xia2021}.

\begin{figure}[t]
  \centering
  \includegraphics[width=\columnwidth]{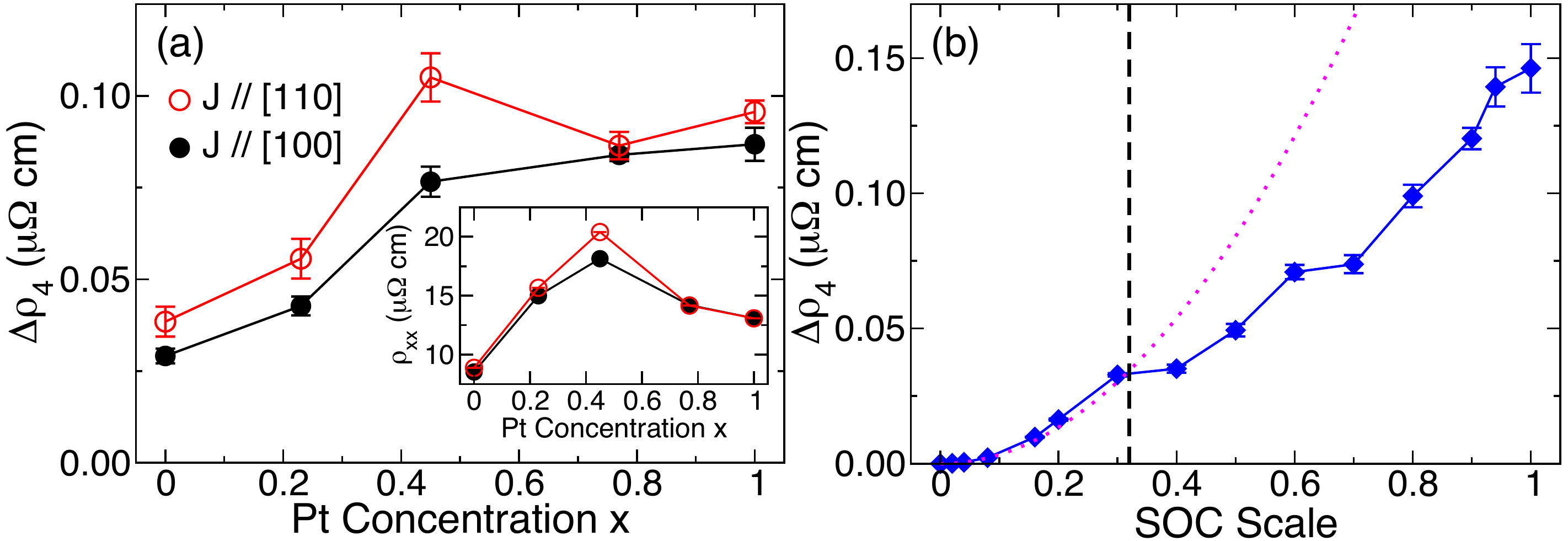}
  \caption{(a) Measured $\Delta\rho_4$ of L1$_0$ ordered Fe$_{0.5}$(Pd$_{1-x}$Pt$_x$)$_{0.5}$ 20-nm-thick films as a function of Pt concentration $x$ at 10~K. The inset shows the measured $\rho_{xx}$. (b) Calculated $\Delta\rho_4$ of FePt as a function of scaled SOC strength. The black dashed line indicates the SOC strength of FePd. The dotted line illustrates quadratic dependence.}
  \label{fig5}
\end{figure}
The isoelectronic properties of Pt and Pd enable the SOC strength to be mediated by changing the Pt/Pd atomic concentration, whereas other parameters, such as the lattice constant, saturation magnetization, and Curie temperature, remain almost the same~\cite{He2012,He2013}. We measure the resistivity and AMR in (001) L1$_0$ Fe$_{0.5}$(Pd$_{1-x}$Pt$_x$)$_{0.5}$ for different Pt concentrations $x$. The sheet resistivity exhibits a maximum near $x=0.5$ and thus obeys Nordheim's rule~\cite{Nordheim1931}, whereas the magnitude of $\Delta\rho_4$ increases with $x$ and is nearly independent of the current orientation as shown in Fig.~\ref{fig5}(a). The enhanced $\Delta\rho_4$ with increasing SOC strength is reproduced by first-principles transport calculation by artificially reducing SOC of FePt; see Fig.~\ref{fig5}(b). At small SOC strength, $\Delta\rho_4$ exhibits a quadratic dependence on SOC.

{\it\color{red}Summary.---} We studied the AMR of (001) L1$_0$ FePt epitaxial films by systematically varying the degree of chemical order and temperature and identified the underlying microscopic mechanisms of AMR. Twofold AMR arises from the competition between the intrinsic mechanism due to magnetization-orientation-dependent band crossings and the extrinsic scattering mechanism that results from thermal and chemical disorder scattering. Fourfold AMR is attributed to the variation in the DOS and hence in the relaxation time at the Fermi surface that is induced by rotating the magnetization. Current-orientation independence is the main criterion used to identify this mechanism. The relaxation time anisotropy is universal for other FMs with proper symmetry and is a possible mechanism for the anisotropic Gilbert damping observed in recent experiments.

\acknowledgements
This study was supported by the National Key R\&D Program of China (Grant No. 2017YFA0303202) and National Natural Science Foundation of China (Grant Nos. 11874283, 51801152, 11774064, 12174028 and 11734004).


\end{document}